%% file: GRBHAWC.tex
\documentclass[a4paper,11pt]{article}
\usepackage{pos}
\usepackage{float}

\title{CONSTRAINTS ON THE VERY HIGH ENERGY GAMMA-RAY EMISSION WITH HAWC}
\ShortTitle{GAMMA-RAY EMISSION WITH HAWC}

\author*[a]{Y. Pérez Araujo}
\author[a]{M. M. González}
\author[a]{N. Fraija}
\forColl{HAWC}

\affiliation[a]{Universidad Nacional Autónoma de México,\\
  Street number 3000, México, México}

\emailAdd{yfperez@astro.unam.mx}
\emailAdd{magda@astro.unam.mx}
\emailAdd{nifraija@astro.unam.mx}

\abstract{Gamma-ray bursts (GRBs) are among the most luminous sources in the universe. The nature of their emission at TeV energies is one of the most relevant open issues related to these events.  The temporal and spectral features inferred from the early and late emissions usually known as prompt and afterglow, respectively, can be interpreted within the 
context of the fireball model. The synchrotron self-Compton process is expected during the afterglow phase. We explain how the theoretical SSC light curves can be compared with  hypothetical upper limit located at $z=0.3$. We show the allowed parameter space of the microphysical parameters and density of the circumburst medium. The most restrictive results are obtained when the SSC process lies in the fast cooling regime.}

\FullConference{37$^{\rm{th}}$ International Cosmic Ray Conference (ICRC 2021)\\
		July 12th -- 23rd, 2021\\
		Online -- Berlin, Germany}

\begin{document}
\maketitle
\section{Introduction}
Gamma-ray bursts (GRBs) are the most luminous gamma-ray transient events in the Universe \cite{2015PhR...561....1K}. GRBs are mainly associated with the core collapse of massive stars or the merger of compact object binaries when the duration of their main emission is longer or less than few seconds, respectively \cite{1998Natur.395..670G, 1992Natur.357..472U, 1999Natur.401..453B, 1992ApJ...392L...9D}. The temporal and spectral features inferred from the early and late emissions usually known as prompt and afterglow, respectively, can be interpreted within the context of the fireball model \citep{1978MNRAS.183..359C}. Within this framework model, the prompt emission is described by internal shocks \citep{1994ApJ...430L..93R, 1997ApJ...490...92K} and magnetic re-connections \citep{2000ApJ...537..810W}, which convert a significant fraction of the kinetic and magnetic energy into radiation, and the afterglow is generated by the deceleration of the outflow in the circumburst medium \cite{2017ApJ...837..116B}.

The Large Area Telescope (LAT) instrument on board the Fermi satellite (Fermi-LAT; \cite{2009ApJ...697.1071A}) has detected high-energy emissions, from hundreds of MeV to a few GeV. These emissions are not consistent with an extrapolation of the prompt emission at keV-MeV energies and come late and has different temporal evolution (eg, \cite{2010PhRvD..82i2004A,2014Sci...343...42A}). Also, 100-400 GeV photons were associated with the afterglow emission observations of the GRB180720B reported by H.E.S.S \cite{2019Natur.575..464A}. Lastly, very-high energy (VHE) photons with energies above 300 GeV were detected from the long GRB 190114C \citep{2019ATel12390....1M} by the MAGIC telescopes for more than 1000 s \cite{2019ATel12390....1M}. 

VHE emission is expected from the nearest and luminous bursts \citep{2020arXiv200311252F,2019ApJ...885...29F}, mainly because of its attenuation with the Extragalactic Background Light. During the afterglow phase, relativistic electrons are accelerated in forward shocks and cooled down by synchrotron and synchrotron-self Compton (SSC) processes \citep{2019ApJ...883..162F}. 

Within the synchrotorn forward shock model, photons from radio wavelengths to gamma-rays are expected, the SSC model provides photons up to the GeV - TeV energy range \citep{2001ApJ...559..110Z}. In this  work, we focus on short GRBs which are closer than long GRBs to the average redshift of 0.48 and, are likely surrounded by a homogeneous interstellar medium \citep{2014ARA&A..52...43B}. We obtain expressions for VHE light curves of the afterglow emission in the SSC model assuming a homogeneous medium. We explain how these light curves can be compared with observed upper limits to restrict the microphysical parameters as in the different cooling phases. We show results for a hypothetical burst with  X-ray fluence of $5\times 10^{-7}\,{\rm erg \,cm^{-2}}$ and an upper limit for the VHE fluence in the energy range of hundreds of GeV of $1\times10^{-6}\,{\rm erg \,cm^{-2}}$. These values were chosen since they are typical for bursts observed by Fermi-GBM and the HAWC observatory, two monitor instruments.

\section{SSC Model}

\begin{figure}[ht!]
\centering
\includegraphics[scale=0.4]{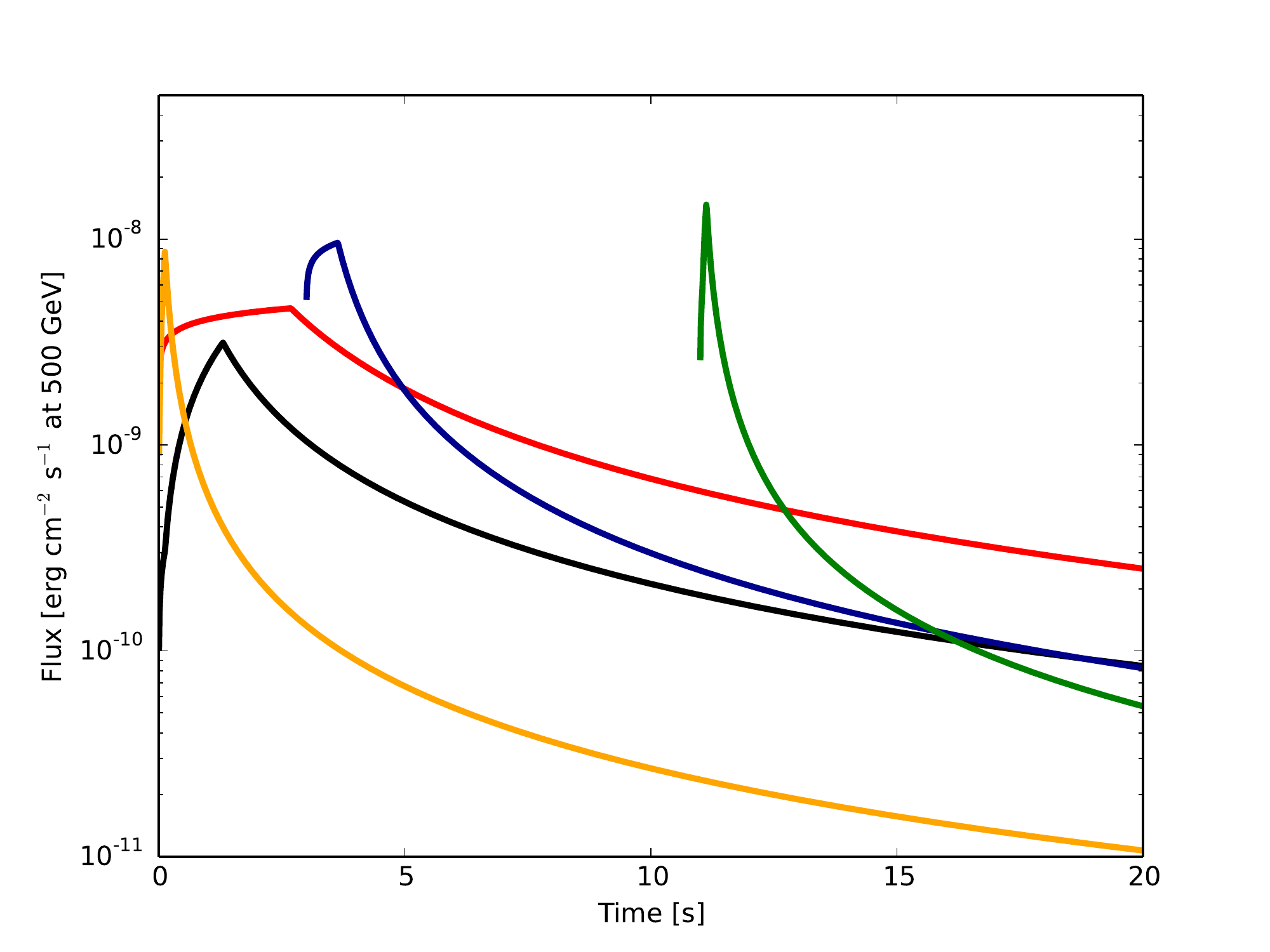}
\caption{For illustrative purposes the flux as a function of time predicted by the SSC model as described in red, blue and green lines show the theoretical light curves in the fast cooling regime assuming different combination of microphysical parameters ([$\epsilon_{B}$=$1.4\times10^{-2}$,$\epsilon_{e}$=$2.6\times10^{-2}$],[$\epsilon_{B}$=$6.5\times10^{-3}$, $\epsilon_{e}$=$1.3\times10^{-2}$] and [$\epsilon_{B}$=$5.7\times10^{-4}$, $\epsilon_{e}$=$7.1\times10^{-3}$], respectively) and different start times ($t_{\rm start}=0$ sec, $t_{\rm start}=3$ seconds and $t_{\rm start}=11$ sec, respectively). Slow cooling regime light curves are plotted in orange and black are derived assuming [$\epsilon_{B}$=$1.9\times10^{-4}$, $\epsilon_{e}$=$8.0\times10^{-3}$] and [$\epsilon_{B}$=$7.8\times10^{-6}$, $\epsilon_{e}$=$4.5\times10^{-2}$], respectively. For all the cases we assume a redshift of $z=0.3$, n = 1 ${\rm cm^{-3}}$ and the isotropic energy of $E_{\rm iso}=3.6\times10^{51}\,{\rm erg}$.
\label{fig:th_lc}}
\end{figure}

We have extended the model presented by \cite{1998ApJ...497L..17S} where the spectrum and light curves for the synchrotron radiation are developed in detailed. The SSC is developed by \cite{2000ApJ...544L..17P} and extended for the slow cooling regime by \cite{2000ApJ...532..286K}. For the SSC scenario, in \cite{2017ICRC...35..620D} we present the computation of the spectral breaks, the maximum flux and the light curves for non-relativistic fast and slow cooling regimes. These calculations assume a photon spectrum described by three power-laws defined by the characteristic (${\rm E_m}$) and cooling (${\rm E_c}$) synchrotron energy breaks and an electron spectral index of ${\rm p=2.4}$. An expression for the energy break ($E_{\rm KN}$) in the Klein-Nishina (KN) regime is also given. The required information to obtain the theoretical light curves is the apparent isotropic kinetic energy of the blast wave ($E_{\rm iso}$), the density of surrounding medium ($n$), the redshift ($z$), the luminosity distance ($D_z$) from the burst to the Earth, the fraction of energy given to the magnetic field (${\rm \epsilon_B}$) and electrons (${\rm \epsilon_e}$).  Figure \ref{fig:th_lc} shows examples of theoretical SSC light curves. As observed, some light curves appear sharp while others are wider in time. The start time is a parameter chosen between 1 and 20 seconds, and together with the density of the surrounding medium and the kinetic energy, define the bulk Lorentz factor. We assume an efficiency of $20\% $ between the kinetic and radiated energy.

\begin{figure}[htp]
\centering
  \includegraphics[clip,width=0.33\columnwidth]{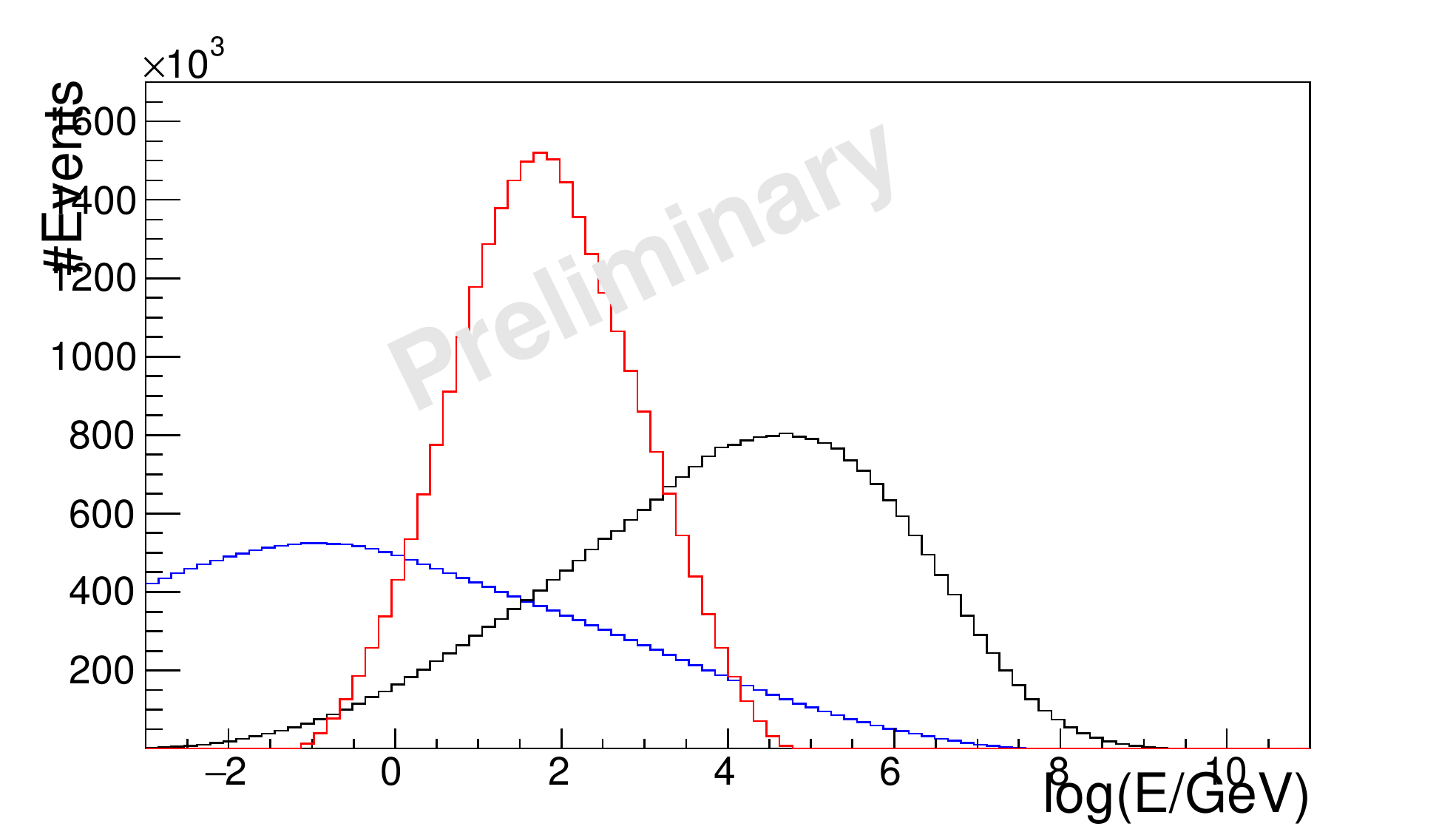}
  \includegraphics[clip,width=0.33\columnwidth]{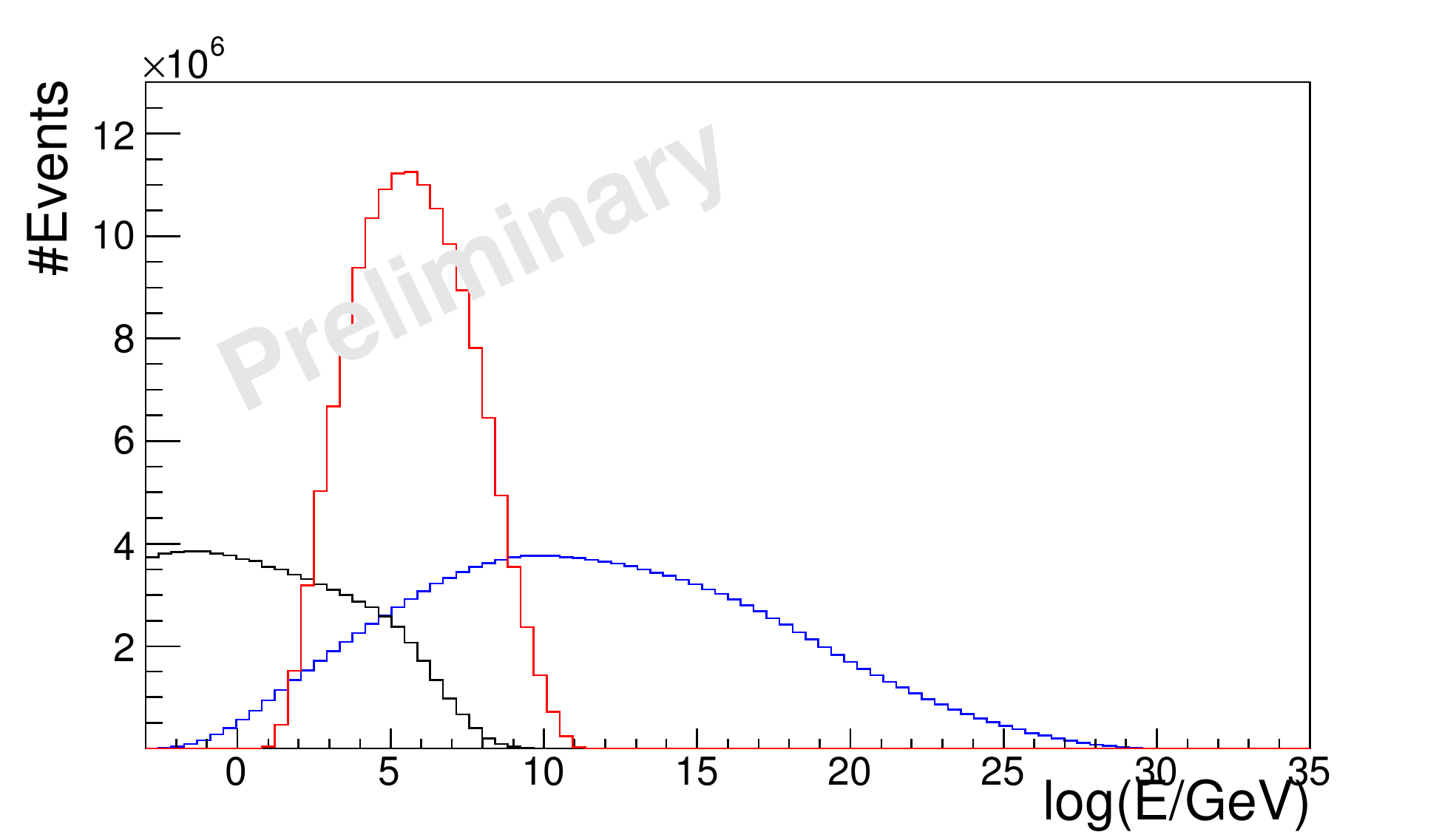}%
  \includegraphics[clip,width=0.33\columnwidth]{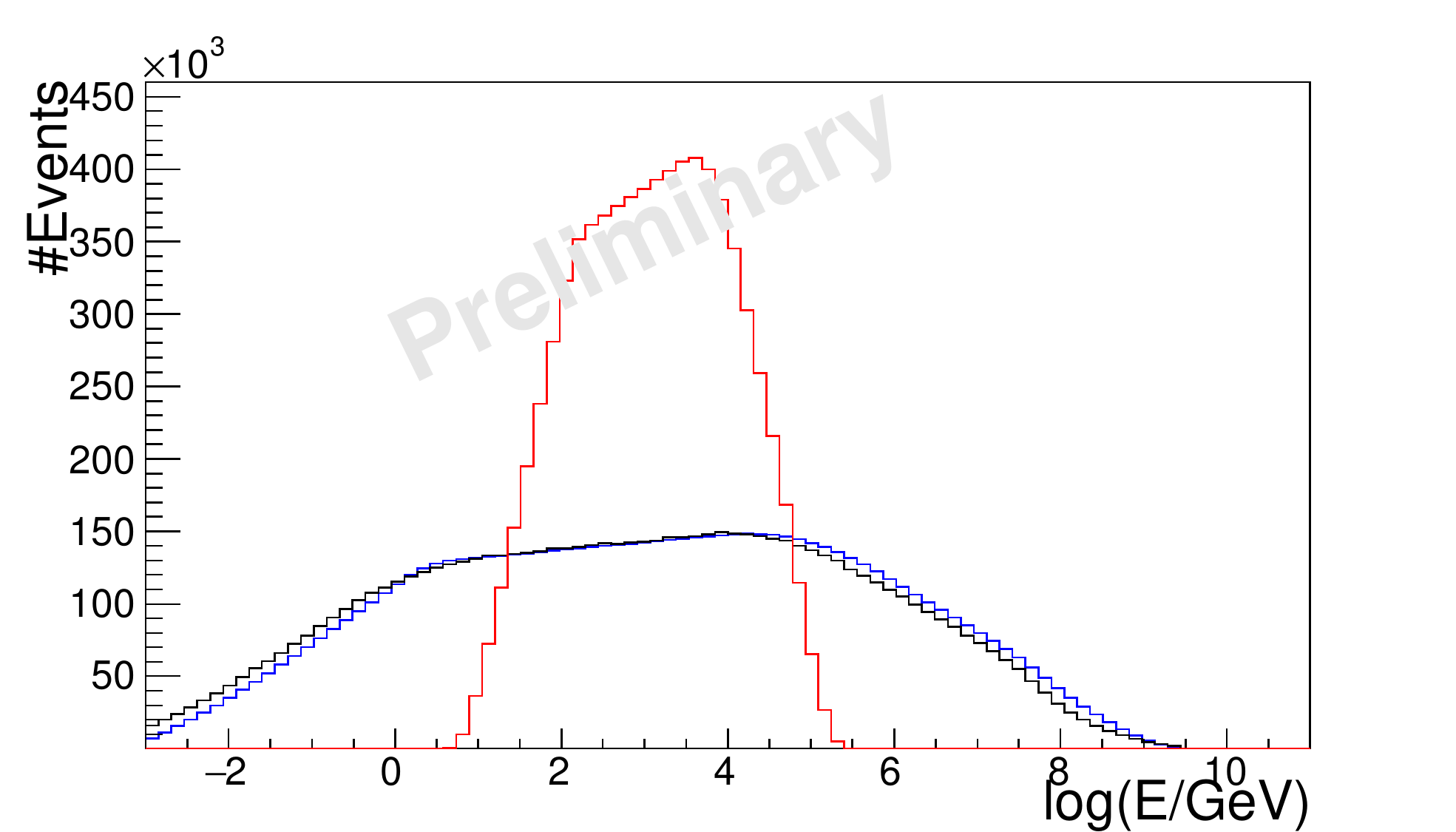}%
\caption{Histogram of SSC energy breaks. Lines in black, blue and red colors show the characteristic, cooling  and Klein-Nishina break when SSC process lies in the fast (left), slow (middle) and the transition (right) for fast (left), slow (middle) and transition (right) cooling cases.
}
\label{fig:EmEcEk}
\end{figure}

For the analysis presented here, we have defined three cases: purely fast cooling, purely fast cooling and the transition regimes. The pure fast cooling regime is defined when ${\rm E_m > E_c}$ from 0 to 20 seconds. The purely slow cooling regime is defined when  ${\rm E_c < E_m}$ from 1 to 20 seconds since the VHE emission from afterglow always gets born as fast cooling regime. finally, the transition regime is defined when the transition from the fast to the slow cooling regime occurs at times later than 1 second and before 20 seconds. Figure ~\ref{fig:EmEcEk} shows a histogram of ${\rm E_m}$, ${\rm E_c}$ and ${\rm E_{\rm KN}}$. We would like to compare these light curves with observations of VHE instruments then, we require $E_{\rm KN} > 1$ TeV and the observation energy equals to 500 GeV. This restriction excludes a quarter of the fast cooling cases, almost none of the slow cooling cases, and half of the transition cases. By comparing the number of cases with $E_{\rm m} > 1$ TeV for fast cooling we can conclude that most of the cases will be in the energy range of ${\rm E_c}< E < {\rm E_m}$. Furthermore,  in the slow cooling regime, for most of the cases, the observation energy is below ${\rm E_c}$ thus, a small number of cases will be in the high-energy power law (${\rm E_c} < E$). The transition cases show similar distributions for ${\rm E_m}$ and ${\rm E_c}$ that crosses at the observation energy.

\section{Analysis and Results}

We calculate the theoretical light curves varying the parameters $\epsilon_{\rm B}$, $\epsilon_{\rm e}$ and $n$ within the ranges of $[10^{-6}, 10^{-1}]$, $[10^{-2}, 10^{-1}]$ and $[10^{-6}, 10^3]\,{\rm cm^{-3}}$, respectively \cite{2014ApJ...785...29S}. To show the potential of the analysis, we have assumed a hypothetical GRB that could be observed by Fermi-GBM and followed up by the HAWC observatory. A typical Fermi-GBM burst in the field of view of HAWC would have an X-ray fluence of $5\times 10^{-7}\,{\rm erg cm^{-2}}$ and a HAWC upper limit for the fluence in the energy range of 80-800 GeV of $1\times 10^{-6}\,{\rm erg cm^{-2}}$ for a short burst as GRB 170206A in a time window of 2 seconds ~\cite{2017ApJ...843...88A}. Then, we have assumed an equal upper limit for ten consecutive time windows from 0 to 20 seconds and compare them to the theoretical light curves at the observation energy of 500 GeV. It is important to mention that in a real case, the observational flux upper limit should be calculated for the corresponding spectral index depending of the cooling regime and its respective power law. We assume $z=0.3$, similar to the average value expected for short GRBs \cite{2014ARA&A..52...43B}.

\begin{figure}[htp]
  \includegraphics[clip,width=0.33\columnwidth]{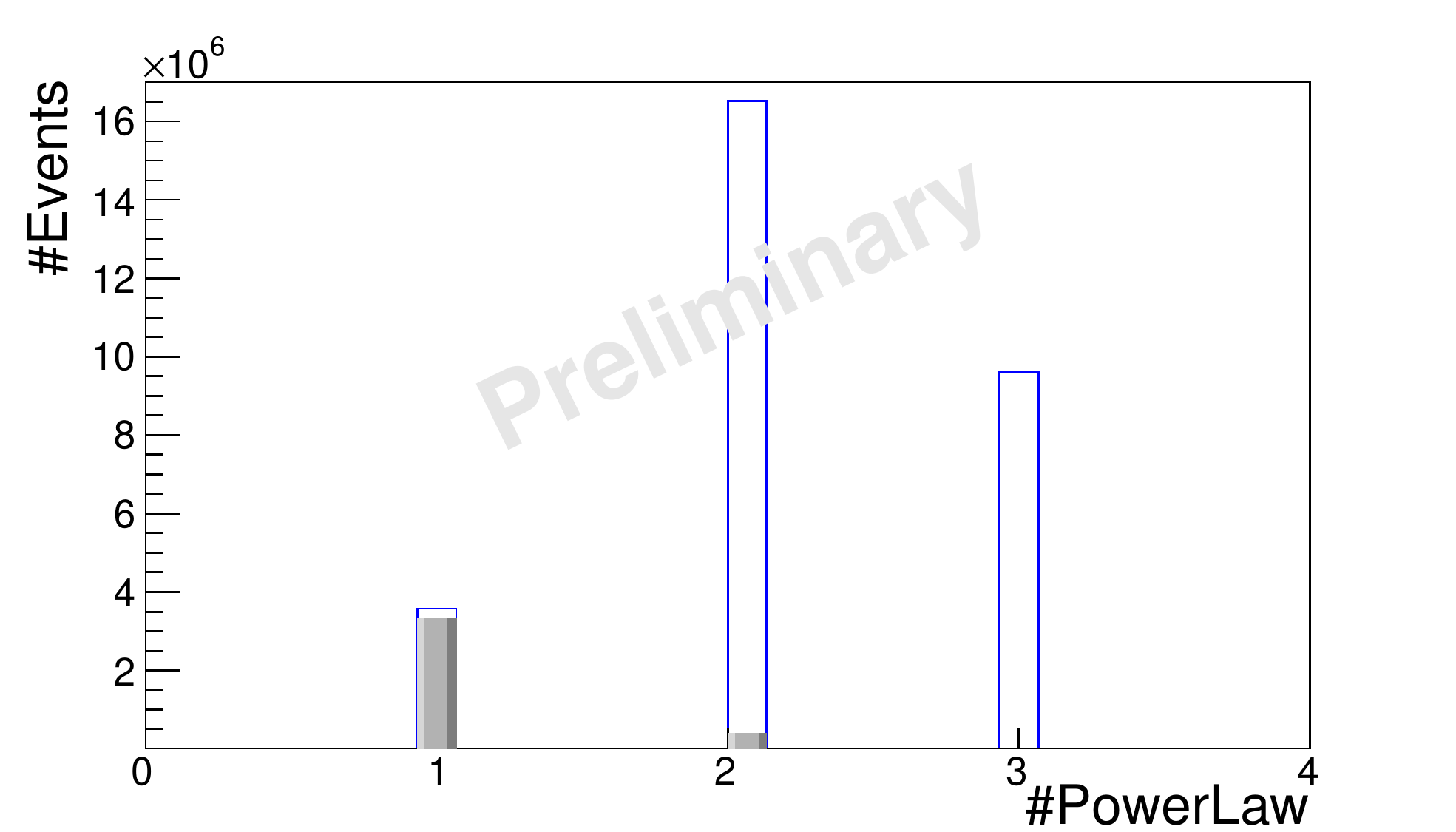}
  \includegraphics[clip,width=0.33\columnwidth]{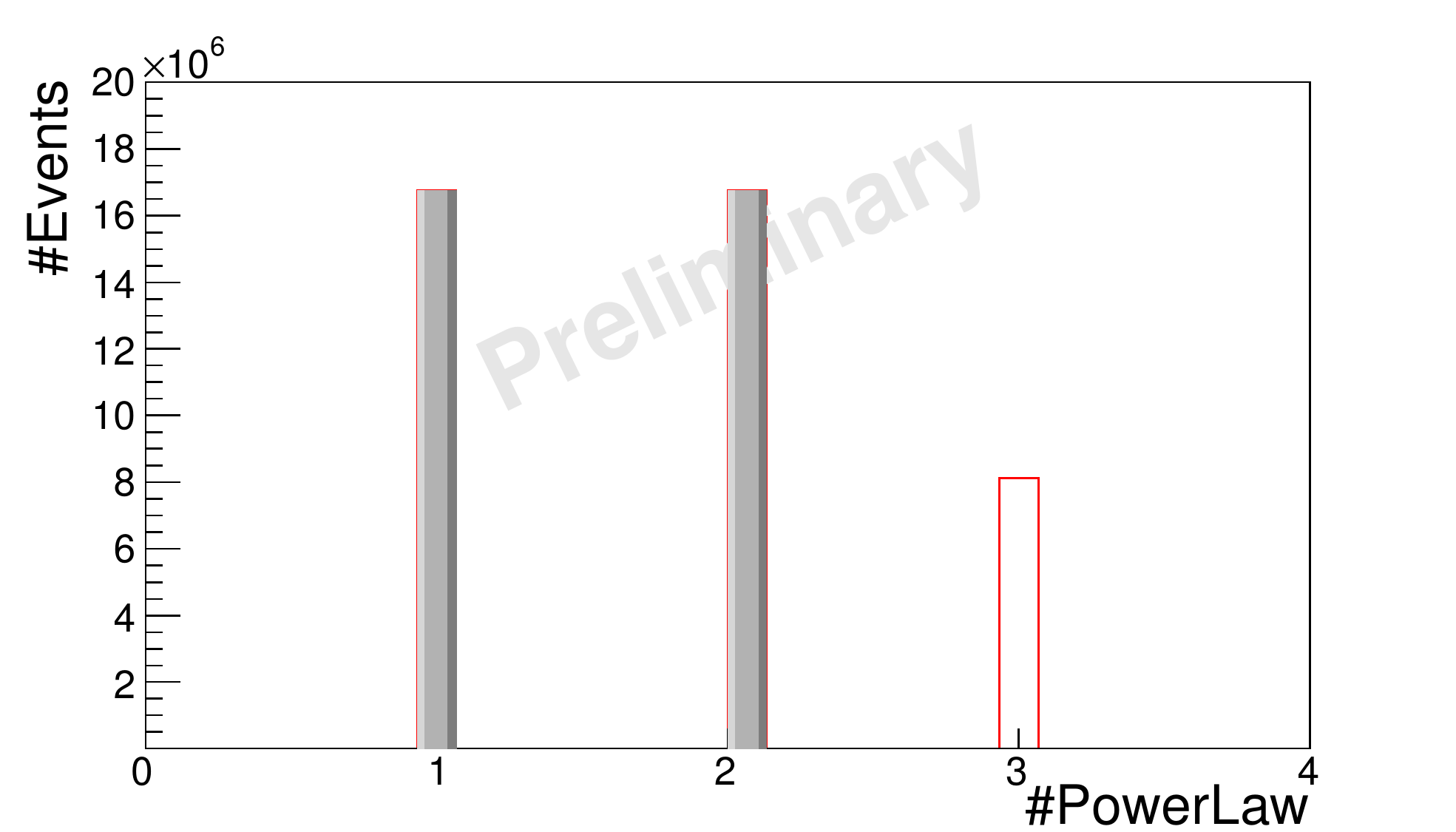}%
  \includegraphics[clip,width=0.33\columnwidth]{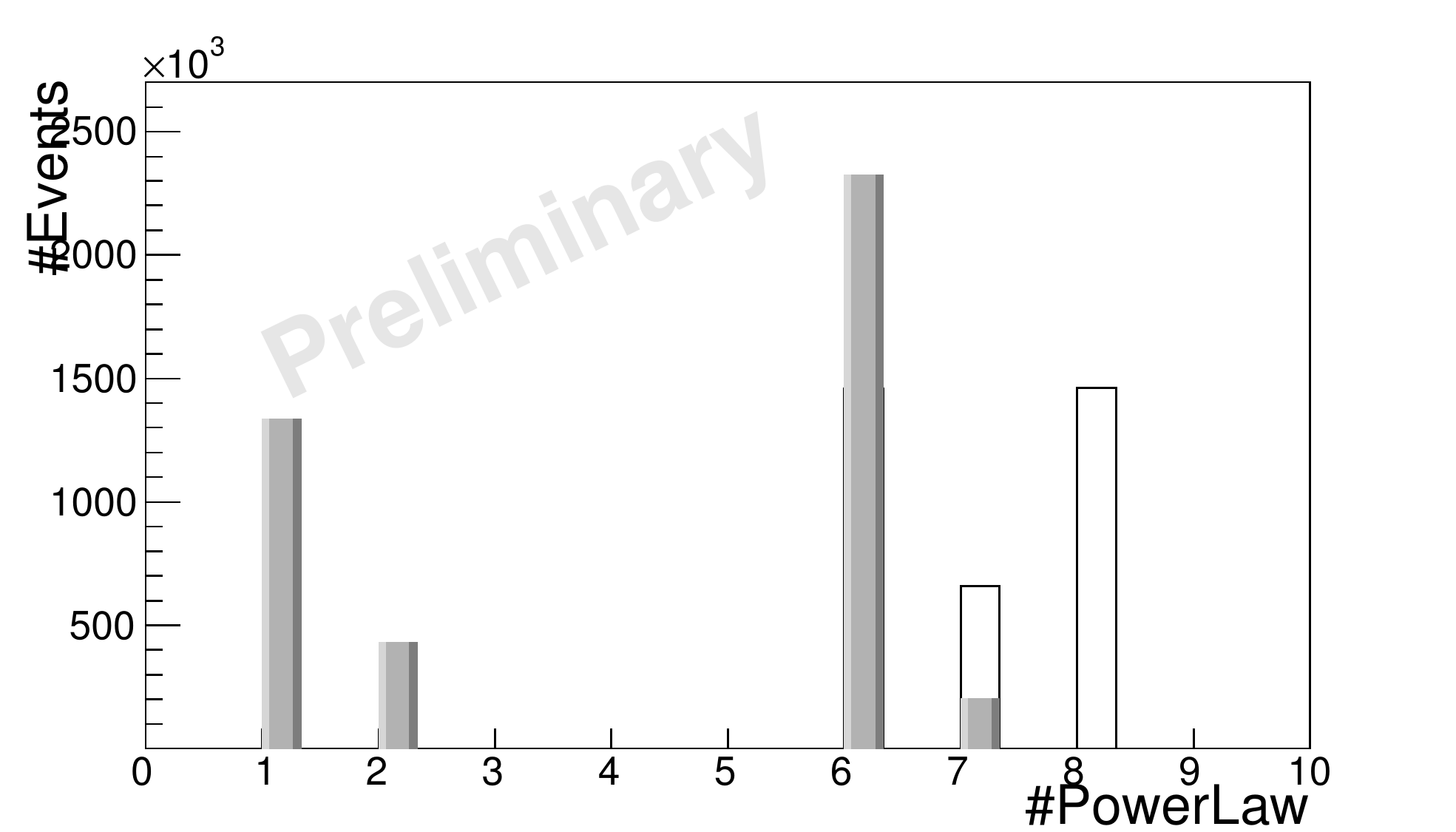}%
\caption{Number of cases (depending of the model parameters) that yield in the lower-energy (values 1 and 6), middle (values 2 and 7) and high-energy (values 3 and 8) power law for each cooling regime. }
\label{fig:PL}
\end{figure}

\begin{figure}[!ht]
\centering
\includegraphics[scale=0.24]{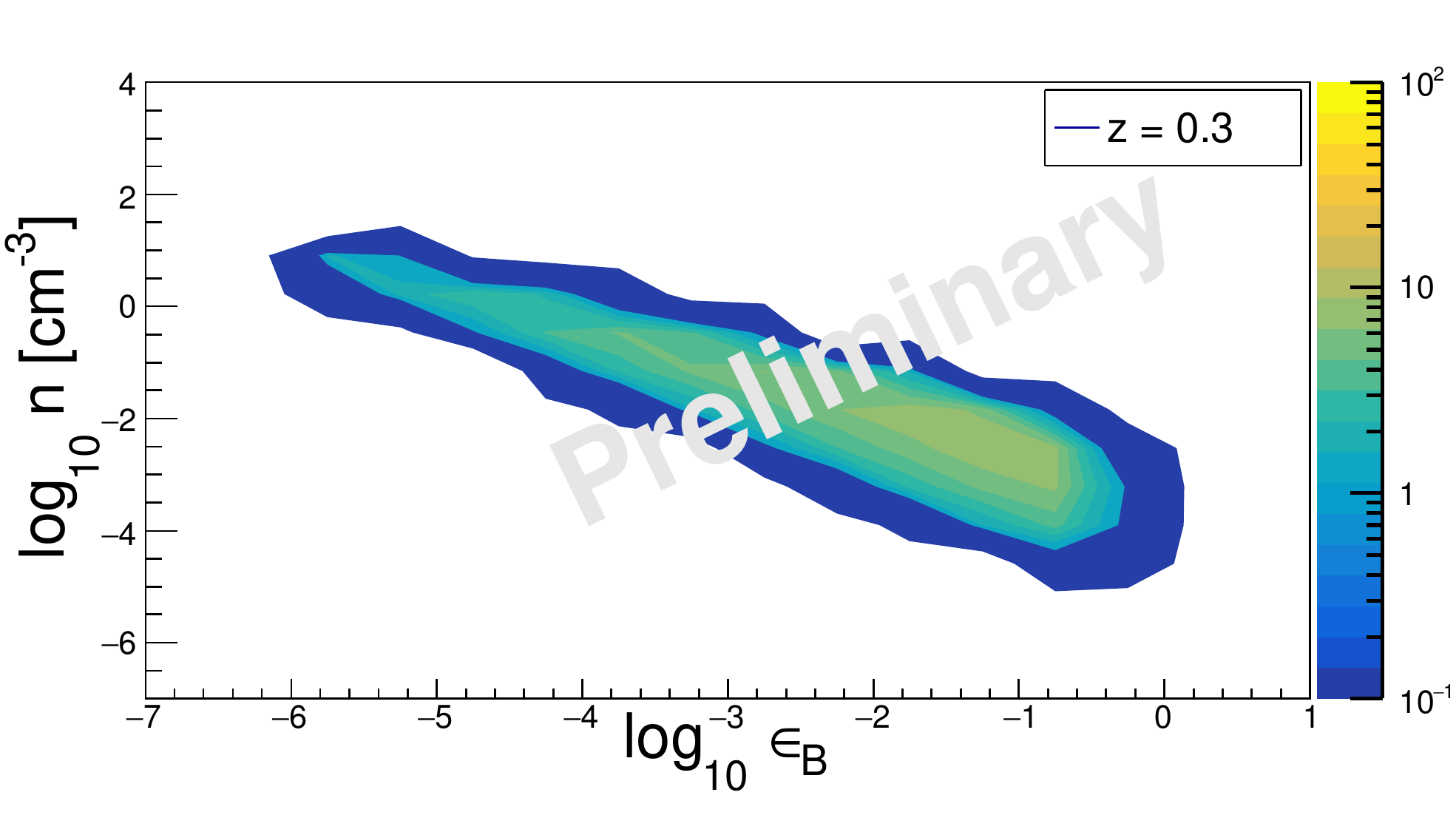}
\includegraphics[scale=0.24]{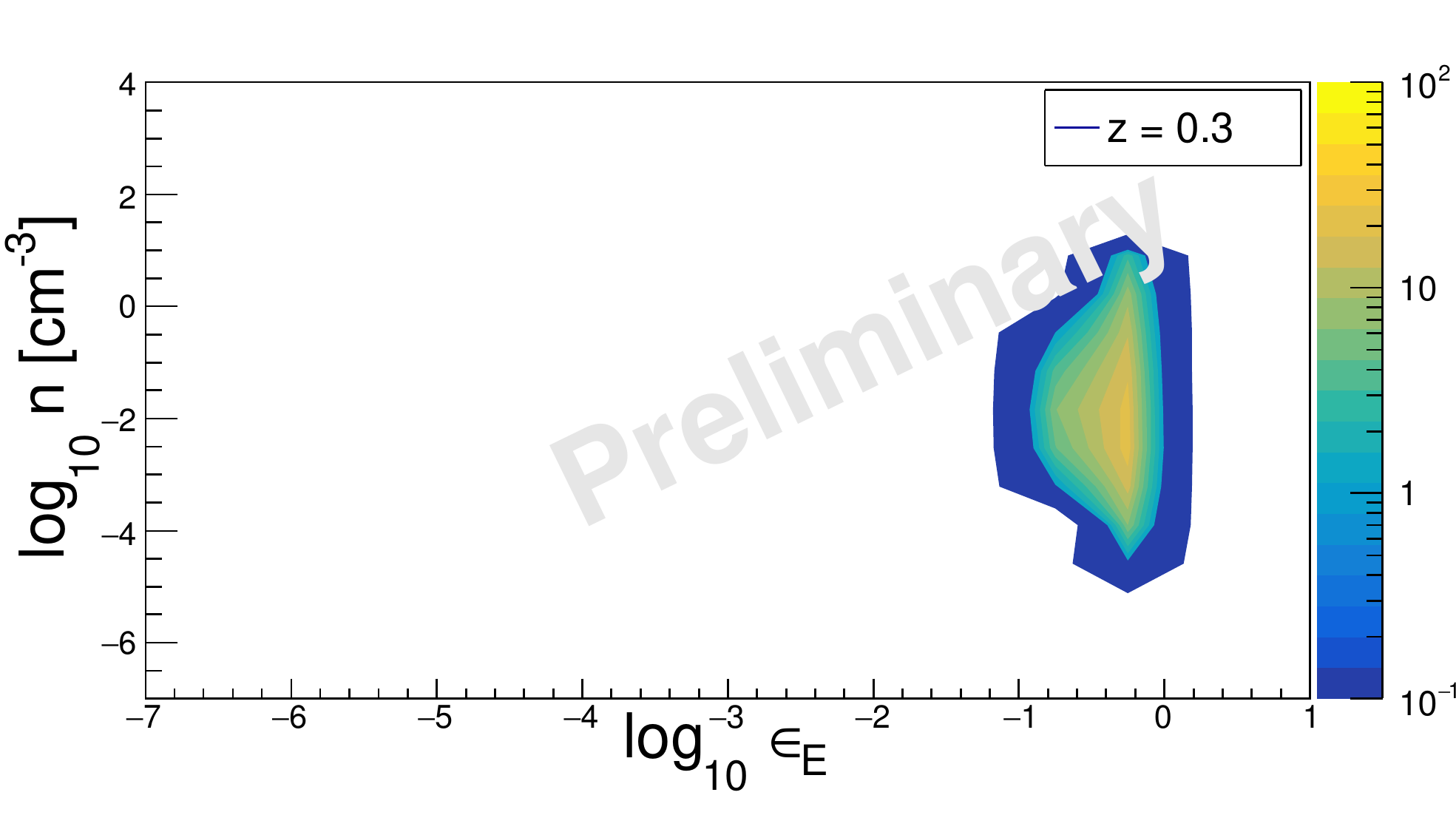}
\includegraphics[scale=0.24]{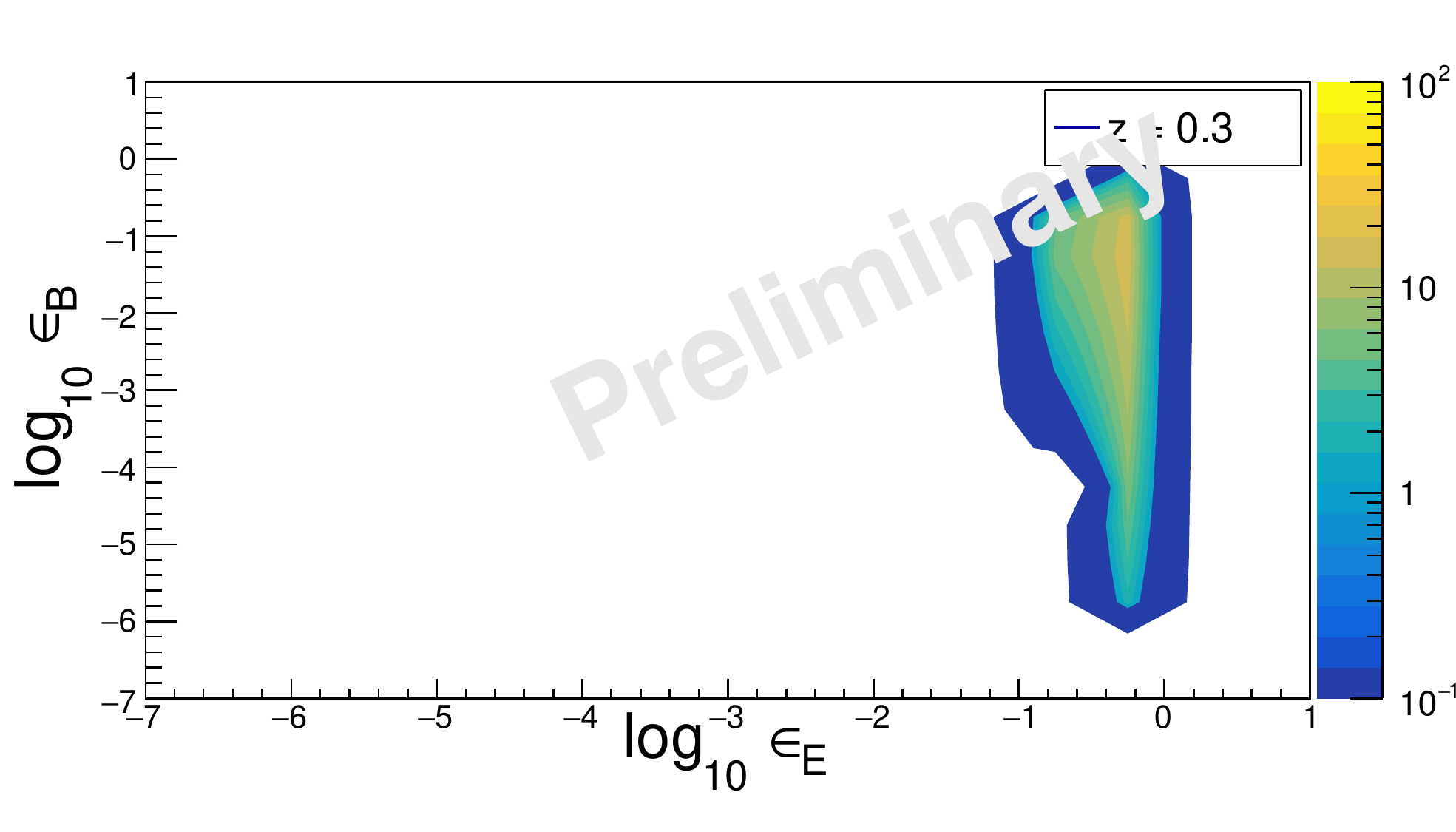}
\caption{Allowed values for microphysical parameter $\epsilon_{B}$,  $\epsilon_{e}$ and density of the external medium when the SSC process lies in a fast cooling regime} \label{fig:PL1}
\end{figure}

\begin{figure}[!ht]
\centering
\includegraphics[scale=0.24]{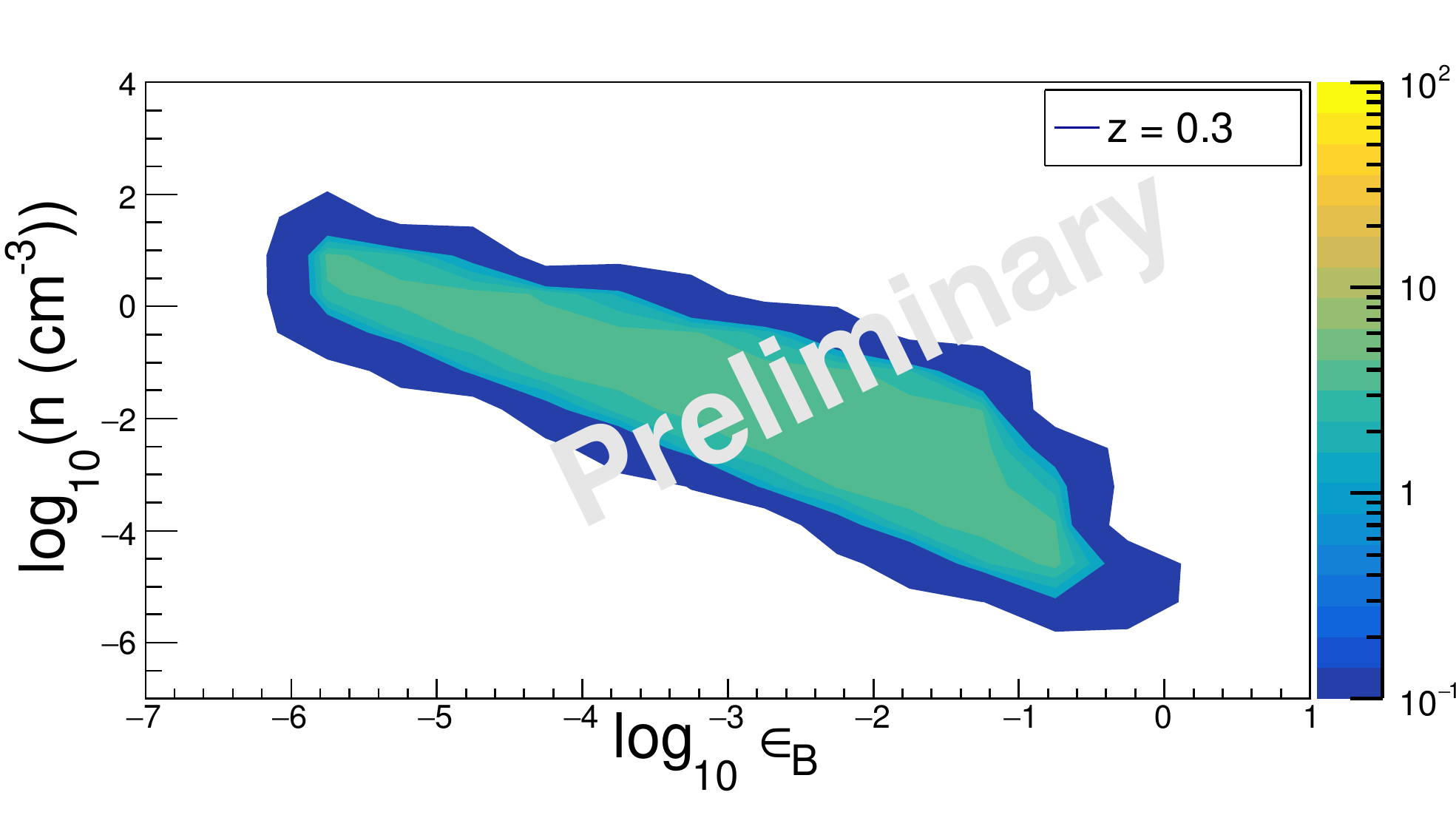}
\includegraphics[scale=0.24]{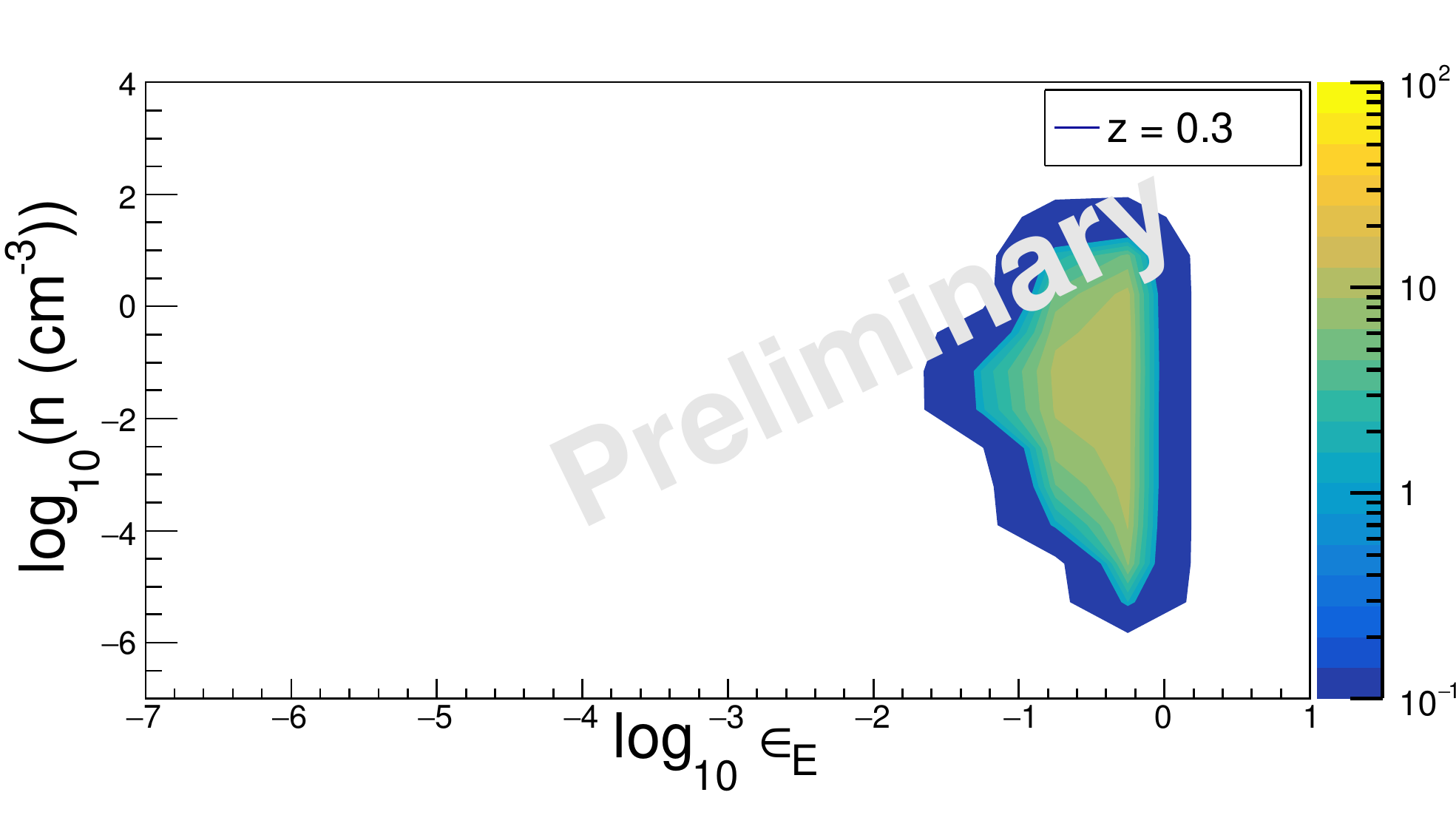}
\includegraphics[scale=0.24]{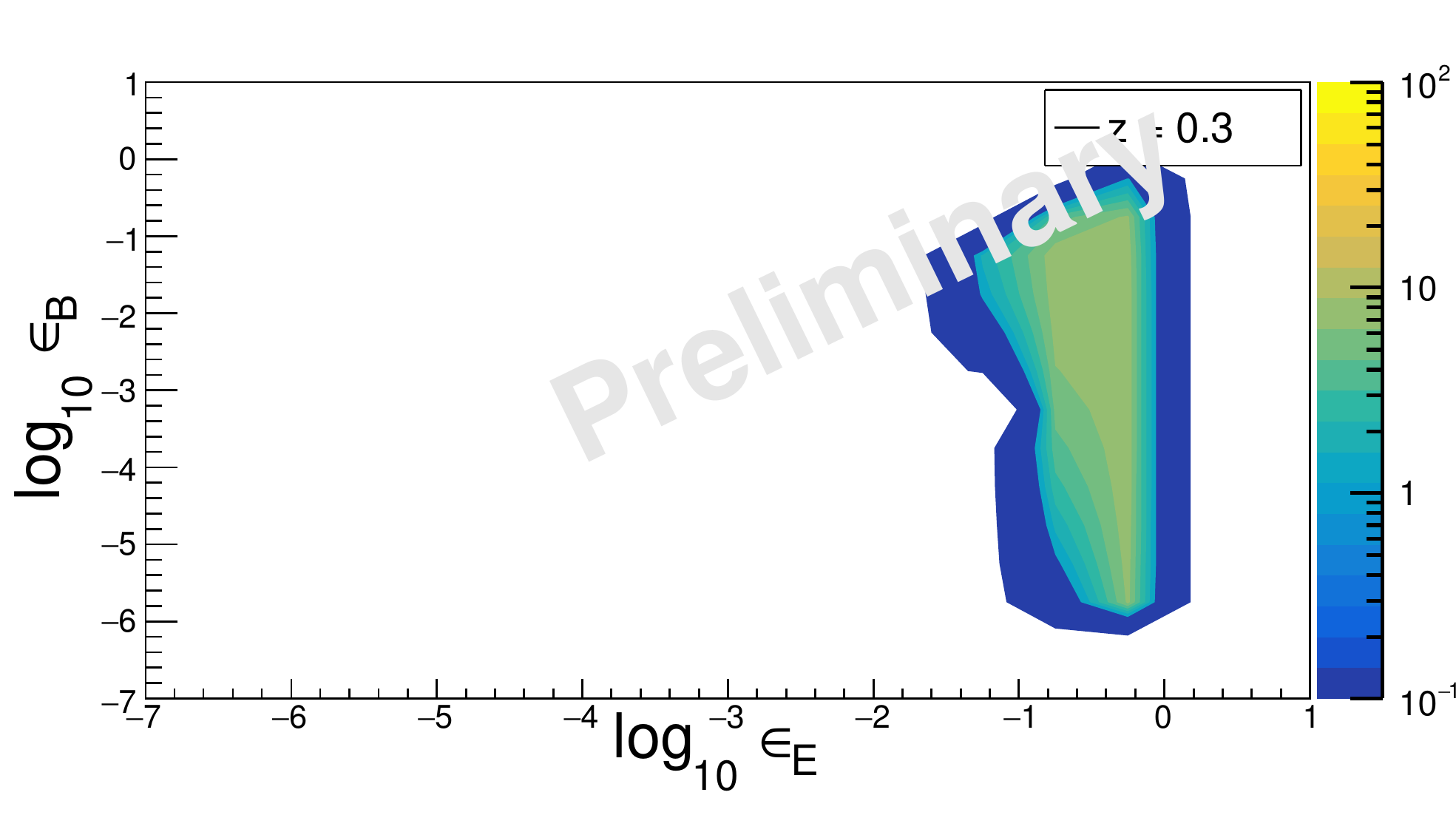}
\caption{The same as Figure 4, but for a transition between fast to slow cooling regime.} \label{fig:PL2}
\end{figure}

\begin{figure}[!ht]
\centering
\includegraphics[scale=0.24]{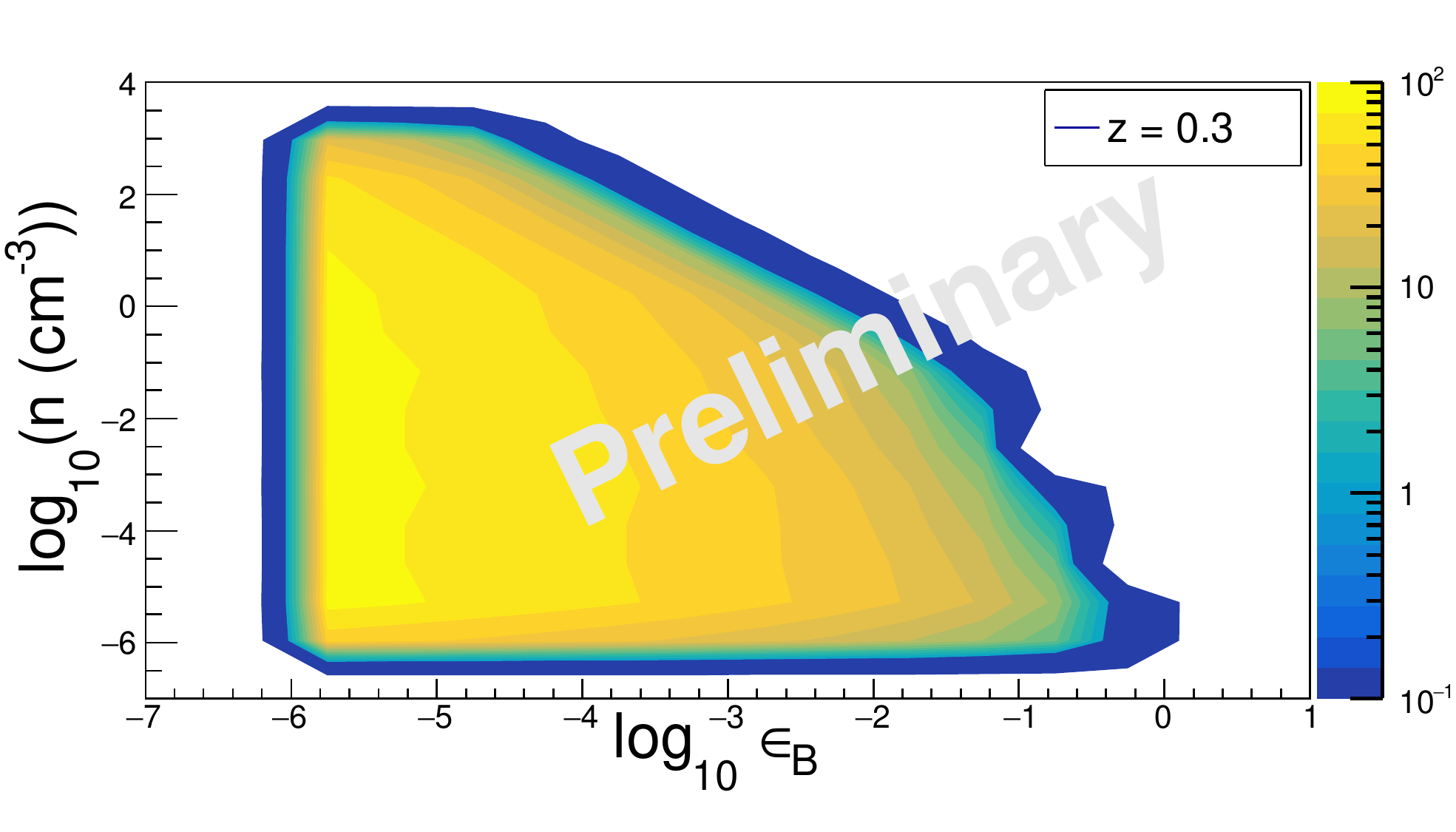}
\includegraphics[scale=0.24]{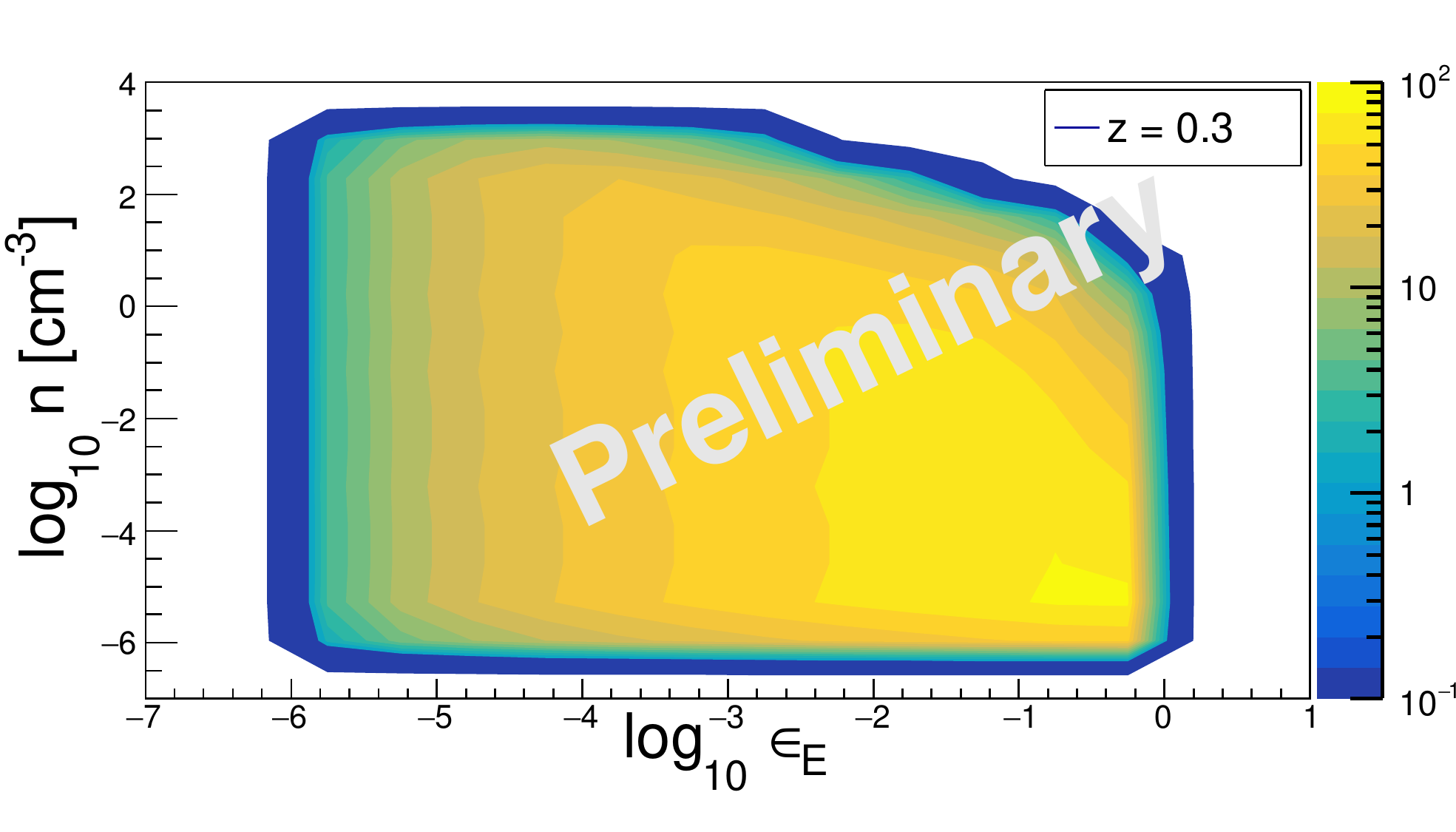}
\includegraphics[scale=0.24]{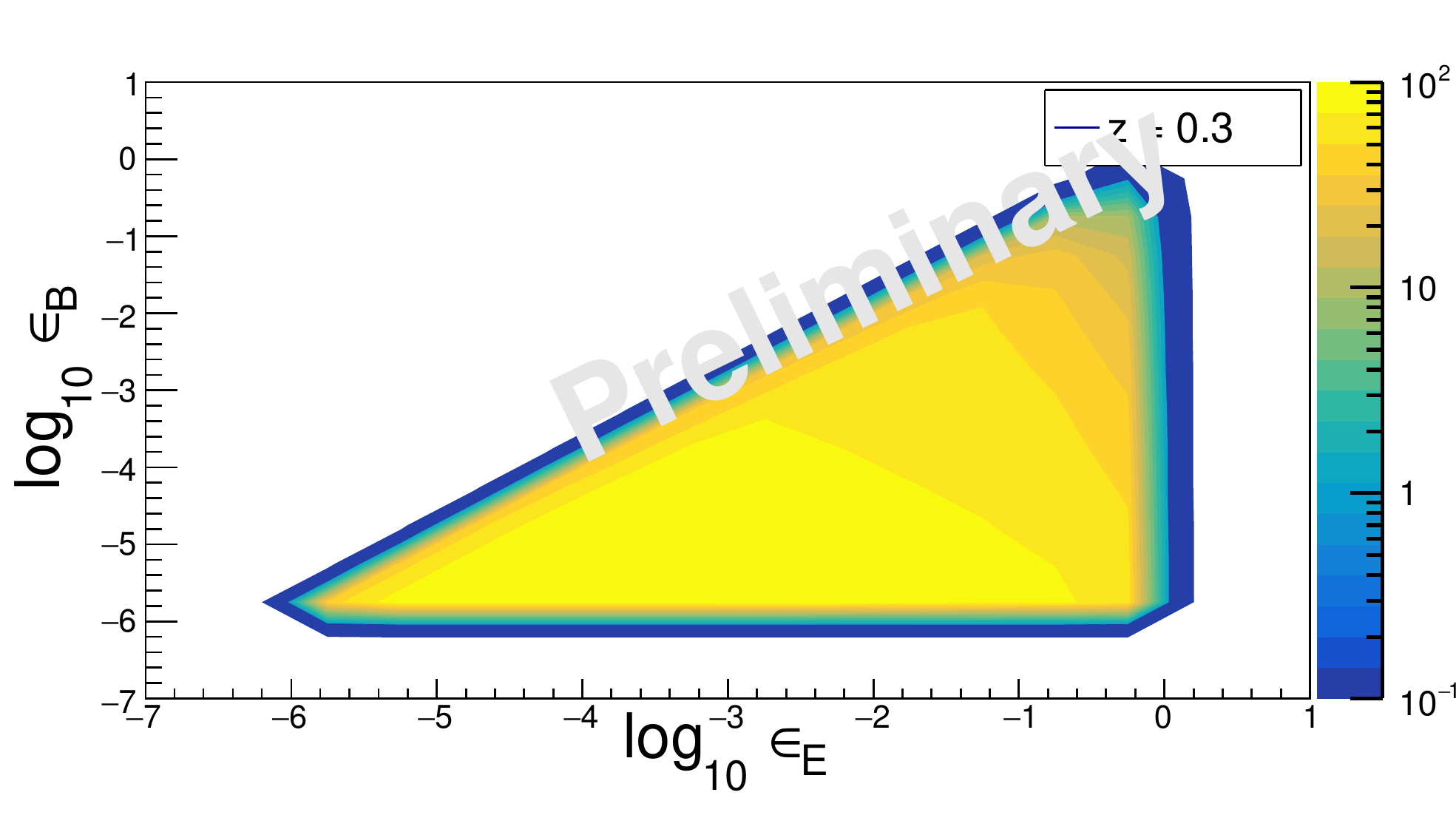}
\caption{The same as Figure 4, but for a slow-cooling regime.} \label{fig:PL3}
\end{figure}

The resulted number of cases in each power law for fast, slow and transition from fast to slow cooling regime is shown in Figure ~\ref{fig:PL}. As observed, the parameter space is mostly restricted for the middle- and high- energy power laws of the fast cooling regime, either in the purely case or the transition case. The allowed values of parameter space for fast, transition and slow cases are shown in Figures~\ref{fig:PL1}, \ref{fig:PL2} and \ref{fig:PL3}. . 


\section{Conclusions}
We have presented the theoretical SSC light curves when the relativistic outflow decelerates in homogeneous circumstellar medium. We have shown the expected light curves when the SSC process lies in the fast, slow and the transition from fast to slow cooling regime.  We have considered a hypothetical GRB located at $z=0.3$ which could have been  detected by Fermi-GBM and followed by the HAWC observatory. We have considered an hypothetical flux upper limit to constrain the microphysical parameters and the circumburst density through a SSC forward shock model.   The  flux upper limit was calculated for the corresponding spectral index  of each  power law and cooling regime. We found that thee parameter space is mostly constrained  for the middle- and high- energy power law of the fast cooling regime, either in the purely (fast or slow) regime or the transition regime.

\acknowledgments
We acknowledge the support from: the US National Science Foundation (NSF); the US Department of Energy Office of High-Energy Physics; the Laboratory Directed Research and Development (LDRD) program of Los Alamos National Laboratory; Consejo Nacional de Ciencia y Tecnolog\'ia (CONACyT), M\'exico, grants 271051, 232656, 260378, 179588, 254964, 258865, 243290, 132197, A1-S-46288, A1-S-22784, c\'atedras 873, 1563, 341, 323, Red HAWC, M\'exico; DGAPA-UNAM grants IG101320, IN111716-3, IN111419, IA102019, IN110621, IN110521; VIEP-BUAP; PIFI 2012, 2013, PROFOCIE 2014, 2015; the University of Wisconsin Alumni Research Foundation; the Institute of Geophysics, Planetary Physics, and Signatures at Los Alamos National Laboratory; Polish Science Centre grant, DEC-2017/27/B/ST9/02272; Coordinaci\'on de la Investigaci\'on Cient\'ifica de la Universidad Michoacana; Royal Society - Newton Advanced Fellowship 180385; Generalitat Valenciana, grant CIDEGENT/2018/034; Chulalongkorn University’s CUniverse (CUAASC) grant; Coordinaci\'on General Acad\'emica e Innovaci\'on (CGAI-UdeG), PRODEP-SEP UDG-CA-499; Institute of Cosmic Ray Research (ICRR), University of Tokyo, H.F. acknowledges support by NASA under award number 80GSFC21M0002. We also acknowledge the significant contributions over many years of Stefan Westerhoff, Gaurang Yodh and Arnulfo Zepeda Dominguez, all deceased members of the HAWC collaboration. Thanks to Scott Delay, Luciano D\'iaz and Eduardo Murrieta for technical support.
\bibliographystyle{JHEP}
\bibliography{references}


\input{authortexv6.tex}

\end{document}

%% file: authortexv6.tex
\clearpage
\section*{Full Authors List: \Coll\ Collaboration}
\scriptsize
\noindent
A.U. Abeysekara$^{48}$,
A. Albert$^{21}$,
R. Alfaro$^{14}$,
C. Alvarez$^{41}$,
J.D. Álvarez$^{40}$,
J.R. Angeles Camacho$^{14}$,
J.C. Arteaga-Velázquez$^{40}$,
K. P. Arunbabu$^{17}$,
D. Avila Rojas$^{14}$,
H.A. Ayala Solares$^{28}$
% ...

\noindent
% ...
$^{14}$Instituto de Física, Universidad Nacional Autónoma de México, Ciudad de México, México,
% ...
$^{17}$Instituto de Geofísica, Universidad Nacional Autónoma de México, Ciudad de México, México,
% ...
$^{21}$Physics Division, Los Alamos National Laboratory, Los Alamos, NM, USA,
% ...
$^{28}$Department of Physics, Pennsylvania State University, University Park, PA, USA,
% ...
$^{40}$Instituto de Física y Matemáticas, Universidad Michoacana de San Nicolás de Hidalgo, Morelia, Michoacán, México,
$^{41}$FCFM-MCTP, Universidad Autónoma de Chiapas, Tuxtla Gutiérrez, Chiapas, México,
% ...
$^{48}$Department of Physics and Astronomy, University of Utah, Salt Lake City, UT, USA,
% ...
